\documentclass[conference,letterpaper]{IEEEtran}
\usepackage{amsmath,amsfonts}
\usepackage{algorithmic}
\usepackage{array}
\usepackage{float}
\usepackage[caption=false,font=normalsize,labelfont=sf,textfont=sf]{subfig}
\usepackage{textcomp}
\usepackage{stfloats}
\usepackage{url}
\usepackage{verbatim}
\usepackage{float}   
\usepackage{algorithm}
\usepackage{graphicx}
\usepackage{cite}
\usepackage{xcolor}
\usepackage{booktabs}
\usepackage{tabularx}

\usepackage{balance}
\usepackage{amssymb,amsthm,amsmath,epsfig,latexsym,graphicx,bm,nccmath}
\allowdisplaybreaks[4] % 0..4, higher = more willing to break
\usepackage{mathtools}

\usepackage[skip=0pt]{caption}
\usepackage[font=footnotesize,skip=0pt]{subcaption}
\begin{document}
\title{\LARGE{Deep Reinforcement Learning for Optimization of STAR-RIS Phase and Energy Splitting Coefficients in OTFS-NOMA Framework}}
\author{
Rais. J. Gachaba\IEEEauthorrefmark{1},
\IEEEauthorblockN{Manobendu Sarker\IEEEauthorrefmark{2}, and Anirban Bhowal\IEEEauthorrefmark{1}}

\IEEEauthorblockA{\IEEEauthorrefmark{1}Department of Electronics and Communication Engineering, National Institute of Technology Rourkela, India}

\IEEEauthorblockA{\IEEEauthorrefmark{2}Poly-Grames Research Center, Department of Electrical Engineering, Polytechnique Montr\'{e}al, Canada}

\IEEEauthorblockA{
    \IEEEauthorrefmark{1}raisgachaba@gmail.com, bhowala@nitrkl.ac.in;
    \IEEEauthorrefmark{2}manobendu.sarker@polymtl.ca
}
}

\maketitle

\begin{abstract}
This paper considers a downlink communication framework comprising a simultaneously transmitting and reflecting reconfigurable intelligent surface (STAR-RIS)-aided by orthogonal time frequency space (OTFS) and non-orthogonal multiple access (NOMA) technologies. Further, delay-Doppler mobility in such frameworks renders classical alternating optimization impractical for per-coherence interval reconfiguration. To mitigate such issues, the STAR-RIS phase-shift and energy-splitting design is formulated as a constrained, non-convex sum-rate maximization problem with closed-form maximum ratio transmission beamforming and fixed NOMA power allocation. To circumvent the per-interval re-optimization burden, a deep reinforcement learning (DRL) approach is adopted that maps observed channel realizations to STAR-RIS configurations through a single forward pass. Specifically, Beta-Space Soft Actor-Critic (SAC-BSE), a maximum entropy DRL agent, is proposed. Simulation results, with two NOMA-multiplexed users on each STAR-RIS branch, confirm rapid convergence, limit the sum-rate degradation to roughly 10\% across a 128-fold user-speed range, and yield consistent gains over OTFS-only, NOMA-only, STAR-RIS-only, fixed-split, and mode-switching baselines as transmit power and the number of STAR-RIS elements increase.
\end{abstract}

% \begin{IEEEkeywords}
% Article submission, IEEE, IEEEtran, journal, \LaTeX, paper, template, typesetting.
% \end{IEEEkeywords}
%\vspace{-2mm}
\section{Introduction}\label{sec:introduction}
Future wireless downlinks (DLs) are expected to serve dense, mobile user populations under stringent spectrum and coverage constraints, which conventional physical-layer designs can address partially only. Non-orthogonal multiple access (NOMA)~\cite{liu_noma_5g} responds to the spectrum-efficiency requirements by enabling multiple users to share the same time-frequency resources through power-domain multiplexing and successive interference cancellation (SIC). Yet, it fails to address the problem of channel time variability. Orthogonal time frequency space (OTFS) modulation~\cite{hadani_otfs,aldababsa_otfs_survey} resolves that gap at the waveform level by multiplexing symbols in the delay-Doppler domain, and it has been shown to pair effectively with NOMA when co-scheduled users exhibit heterogeneous mobility profiles~\cite{ding_otfs_noma}. However, none of the aforementioned schemes addresses the propagation environment itself, which can be deemed as programmable by reconfigurable intelligent surfaces (RIS)~\cite{mohjazi_ai_sdm}. Further, simultaneously transmitting and reflecting RIS (STAR-RIS) can resolve the limitation of RIS by ensuring full-space coverage due to its property of simultaneously reflecting and transmitting the incident signal.

Thus, the amalgamation of multiple technologies such as OTFS, NOMA, and STAR-RIS is a natural fit for high-mobility DL networks. Despite deploying these technologies, tracking the delay-Doppler dynamics of a time-varying OTFS channel requires the STAR-RIS phase shifts and energy-splitting ratios to be reconfigured every coherence interval, which is impractical to be performed using classical alternate optimization (AO) techniques for real-time deployment.

Existing works within this design space reveals noteworthy research problems. Specifically, the rate and phase optimization structure was developed for a single-branch RIS-aided OTFS-NOMA system in ~\cite{li_ris_otfs_noma}, and the deep reinforcement learning (DRL)-based cascaded-channel beamforming formulation was developed for RIS-aided system without OTFS or NOMA in \cite{saglam_ris_miso_pda}. Neither treats the reflection and transmission energy-splitting trade-off that appears when the surface is generalized to a STAR-RIS. This trade-off, namely a STAR-RIS-specific decision variable that must be learned jointly with the phase shifts under a time-varying delay-Doppler channel, is a critical research gap which this paper seeks to address.

Building on this, the problem formulation involves the joint STAR-RIS phase-shift and energy splitting design of a STAR-RIS-aided OTFS-NOMA DL as a constrained, non-convex sum-rate maximization problem. It involves constraints such as user quality-of-service (QoS), per-element lossless energy conservation, and base station transmit power. Unlike prior STAR-RIS-aided OTFS-NOMA formulations that treat the reflection and transmission branches independently, the reflection transmission energy coupling is retained as the residual structure for resolution by a learning-based solution. Further, the work highlights the utilization of closed-form maximum ratio transmission (MRT) beamforming and fixed NOMA weights to isolate the STAR-RIS coefficients as the sole decision variables, leaving the joint two-branch phase and energy-splitting design as the residual optimization structure to be resolved.

Such optimization problem is solved by a DRL approach that maps observed channel realizations to STAR-RIS configurations through a single forward pass, bypassing the per-interval iteration budget that classical AO cannot meet under OTFS mobility. DRL-based designs have already been applied to RIS-assisted ultra reliable and low latency communication (URLLC)~\cite{hashemi_ris_miso_urllc}, aerial NOMA~\cite{umer_aerial_ris}, and satellite~\cite{bao_ris_satellite} systems, motivating its adoption here. Among the available algorithmic choices, we build on the Soft Actor-Critic (SAC) algorithm~\cite{haarnoja_sac}, whose Gaussian policy handles the continuous phase-shift action space without discretization and whose regular exploration of entropy prevents the policy from collapsing into a stale configuration with the evolution of the OTFS channel. Thus, the contributions of this paper can be summarized as follows.
\begin{enumerate}
\item Formulation of STAR-RIS phase-shift and energy-splitting design for a STAR-RIS-aided OTFS-NOMA system as a constrained, non-convex sum-rate maximization problem.
%extending the RIS-aided OTFS-NOMA rate structure of~\cite{li_ris_otfs_noma} and the cascaded-channel formalism of~\cite{saglam_ris_miso_pda} from a single-branch RIS to a two-branch STAR-RIS coupled through per-element energy conservation, wherein closed-form MRT beamforming and fixed NOMA weights isolate the STAR-RIS coefficients as the sole decision variables.
\item A maximum-entropy DRL agent, namely Beta-Space Soft Actor-Critic (SAC-BSE), is developed to decouple the design across sub-networks, wherein an entropy-regularized policy learns the phase manifold while a dedicated Beta-Space Explorer (BSE) sub-network resolves the element-wise energy split under delayed channel state information (CSI).
\item The performance of SAC-BSE is evaluated against baselines such as OTFS-only, NOMA-only, STAR-RIS-only, fixed energy-splitting, and mode-switching DL systems.
%revealing state-dependent gains over every fixed and mode-switching baseline, single-digit sum-rate degradation across a two-orders-of-magnitude user-speed range, and sustained passive beamforming gain as the surface grows.
\end{enumerate}

%%============================================================%%
\vspace{-2.4mm}
\section{System Model}
\label{sec:system_model}

\subsection{Network Topology and STAR-RIS Model}
\label{subsec:topology}

\begin{figure}[tb]
    \centering
    \includegraphics[scale=0.3]{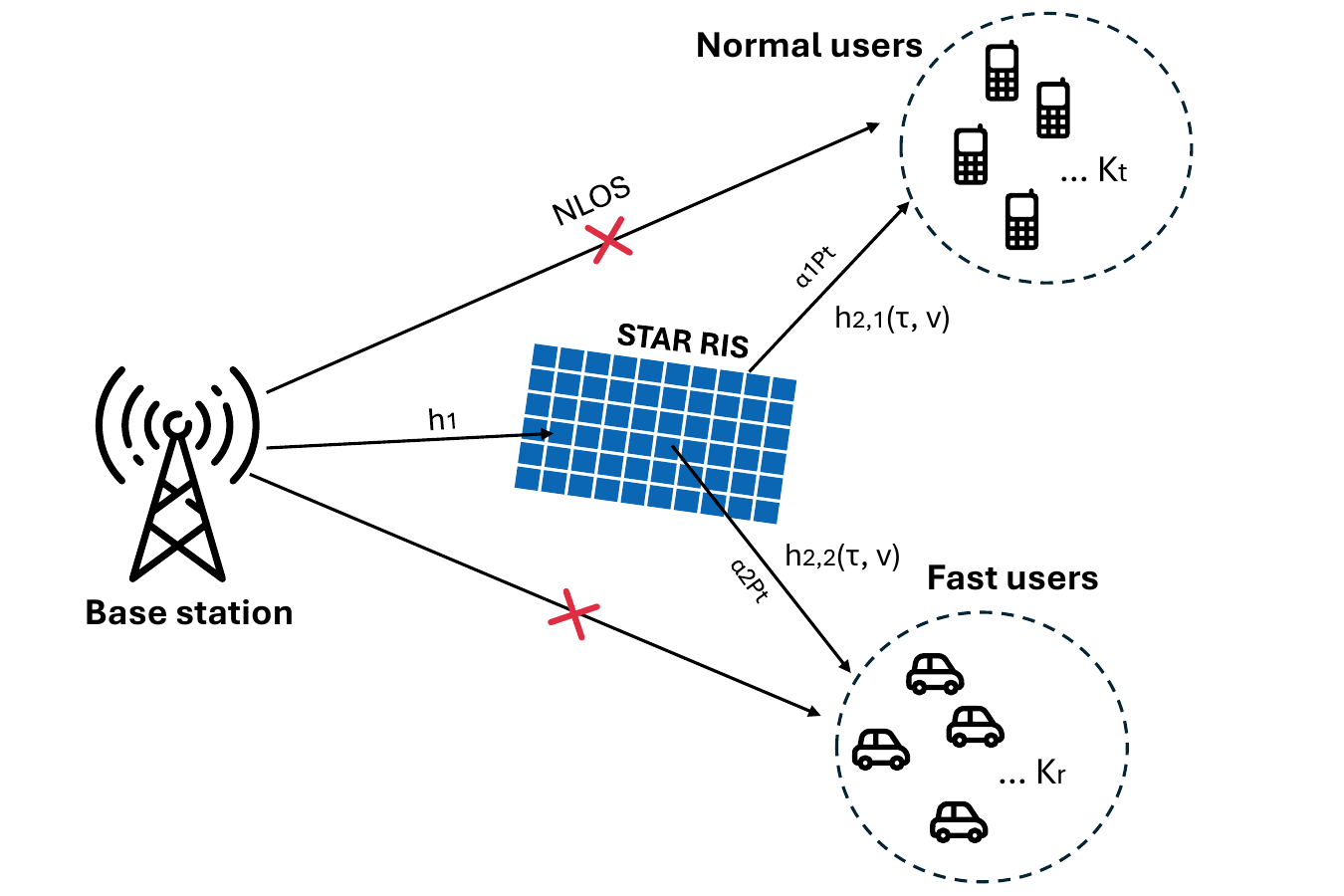}
    \caption{STAR-RIS-aided OTFS-NOMA communication system.}
    \label{fig:star-ris-system-model}
\end{figure}

We consider a DL OTFS-NOMA system in which a base station (BS) equipped with $M$ antennas serves $K$ single-antenna users through a STAR-RIS composed of $L$ reconfigurable elements, as illustrated in Fig.~\ref{fig:star-ris-system-model}. The direct BS-to-user link is assumed to be blocked, such that all DL transmission is routed through the surface. The STAR-RIS operates in energy-splitting (ES) mode, in which each element $l$ simultaneously reflects and transmits the incident signal with coefficients
\begin{equation}
\varphi_{{\rm r},l} = \sqrt{\beta_{\rm r}(l)}\, e^{j\theta_{\rm r}(l)}, \quad \varphi_{{\rm t},l} = \sqrt{\beta_{\rm t}(l)}\, e^{j\theta_{\rm t}(l)},
\end{equation}
where $\beta_{\rm r}(l),\beta_{\rm t}(l)\in[0,1]$ denote the fractions of incident power reflected and transmitted by element $l$, and $\theta_{\rm r}(l),\theta_{\rm t}(l)\in[0,2\pi)$ are the corresponding reflection and transmission phase shifts. Under the standard lossless assumption, energy conservation at each element requires
\begin{equation}
\beta_{\rm r}(l) + \beta_{\rm t}(l) = 1, \quad l=1,\dots,L.
\label{eq:energy_conservation}
\end{equation}

The $K$ users are partitioned into a reflection-side group $U_{\rm r}=\{1,\dots,K_{\rm r}\}$ and a transmission-side group $U_{\rm t}=\{1,\dots,K_{\rm t}\}$, with $K=K_{\rm r}+K_{\rm t}$. Each user $u$ is served through the corresponding STAR-RIS diagonal matrix $\mathbf{\Phi}_u\in\{\mathbf{\Phi}_{\rm r},\mathbf{\Phi}_{\rm t}\}$, with the corresponding diagonal elements being denoted as $\phi_{\rm r}^l$ and $\phi_{\rm t}^{l}$, respectively. As shown in Fig.~\ref{fig:star-ris-system-model}, $U_{\rm r}$ corresponds to the normal mobility users while $U_{\rm t}$ denotes the high mobility (fast) users, such that the transmission side branch is dedicated to the high speed group. Within each branch $b\in\{r,t\}$, co-branch users are further multiplexed in the power domain via NOMA.
% \vspace{-2mm}
\subsection{OTFS Modulation}
\label{subsec:otfs}

For each user $u$, let $x_u[k,l]$, with $k=0,\dots,N-1$ and $l=0,\dots,M_c-1$, denote the delay-Doppler (DD) domain data symbols multiplexed on a uniform DD grid with $N$ Doppler bins and $M_c$ delay bins, following the OTFS framework of~\cite{hadani_otfs}. The DD grid is defined as $\Gamma=\{(k/(NT),\,l/(M_c\Delta f))\}$, where $M_c\Delta f$ is the system bandwidth, $NT$ is one OTFS frame duration, and $\Delta f = 1/T$. Each user's DD-domain symbols are mapped to the time-frequency (TF) domain through an inverse Symplectic Finite Fourier transform (ISFFT)~\cite{raviteja_otfs_io},
\begin{equation}
X_u[n,m]=\frac{1}{\sqrt{NM_c}}\sum_{k=0}^{N-1}\sum_{l=0}^{M_c-1}x_u[k,l]\,e^{j2\pi\left(\frac{nk}{N}-\frac{ml}{M_c}\right)},
\label{eq:isfft}
\end{equation}
for $n=0,\dots,N-1$ and $m=0,\dots,M_c-1$. The TF-domain signal of each user is then precoded by a unit-norm beamforming vector $g_u\in\mathbb{C}^{M\times1}$, $\|g_u\|=1$, and scaled by its NOMA power-allocation coefficient $\eta_u$. Applying the Heisenberg transform yields the spatio-temporal transmit signal
\begin{align}
x(t)=&\sum_{u=1}^{K} g_u \sqrt{\eta_u P_{\rm t}}\sum_{n=0}^{N-1}\sum_{m=0}^{M_c-1}X_u[n,m]\nonumber\\
& \qquad \qquad \times p_{\rm tx}(t-nT)\,e^{j2\pi m\Delta f(t-nT)},
\label{eq:heisenberg}
\end{align}
where $p_{\rm tx}(t)$ is the transmit pulse and $P_{\rm t}$ is the maximum BS transmit power. The NOMA power budget within each branch $b\in\{\rm r,\rm t\}$ is enforced by $\sum_{u:\,\mathrm{branch}(u)=b}\eta_u \leq 1$.
% \vspace{-2mm}
\subsection{Cascaded BS--STAR-RIS--User Channel}
\label{subsec:cascaded_channel}

The BS-to-STAR-RIS link $\mathbf{H}_1\in\mathbb{C}^{L\times M}$ is modeled as quasi-static Rayleigh fading, consistent with the short-range, low-mobility deployment path between the BS and the surface. The STAR-RIS-to-user link, in contrast, must capture the multipath and Doppler dynamics experienced by a mobile user, and is modeled in the DD domain as
\begin{equation}
h_u(\tau,\nu)=\sum_{p=1}^{P_u}h_{u,p}\,\delta(\tau-\tau_{u,p})\,\delta(\nu-\nu_{u,p}),
\label{eq:dd_channel}
\end{equation}
where $P_u$ denotes the number of DD paths between the STAR-RIS and user $u$, and $h_{u,p}$, $\tau_{u,p}$, $\nu_{u,p}$ are the fading coefficient, delay, and Doppler shift of the $p$-th path. The OTFS receiver chain proceeds by first applying a Wigner transform to the TF domain and then an SFFT back to the DD domain. Combining this chain with the cascaded-channel formalism developed for RIS-aided Multiple-input and single-output (MISO) systems in~\cite{saglam_ris_miso_pda}, the effective end-to-end channel of user $u$ can be expressed as $h_{\mathrm{eff},u}^{\rm H} \triangleq h_{u,p}^{\rm H}\,\mathbf{\Phi}_{u}\,\mathbf{H}_1 \in \mathbb{C}^{1\times M}$, such that the DD-domain received signal of user $u$ is given by
\begin{equation}
y_u = h_{\mathrm{eff},u}^{\rm H} x + w_u, \quad w_u \sim \mathcal{CN}(0,\sigma_w^2).
\end{equation}
% \vspace{-3mm}
\subsection{Beamforming, Power Allocation, and Decoding Order}
\label{subsec:beamforming_decoding}

Throughout this work, the transmit beamformer for user $u$ is obtained in closed form via MRT on the estimated effective channel, $g_u = \frac{\hat{h}_{\mathrm{eff},u}}{\|\hat{h}_{\mathrm{eff},u}\|}$, such that $\|g_u\|=1$ and the BS transmit-power budget $\mathrm{tr}(\mathbf{G}\mathbf{G}^{\rm H})\leq P_{\rm t}$ is automatically satisfied by the power-scaled construction in \eqref{eq:heisenberg}. The NOMA power allocation is likewise fixed as a system design parameter, with equal weights $\eta_u = 1/K$ across all users; consequently, the SIC decoding order is determined entirely by the effective channel gain rather than by any power-based fairness weighting.

Within each branch, users perform SIC in ascending order of the effective channel gain $\gamma_u \triangleq |h_{\mathrm{eff},u}^{\rm H}g_u|^2$. A user first removes the signals of all weaker co-branch users, then decodes its own symbol, and treats stronger co-branch users' signals as residual interference.
% \vspace{-2mm}
\subsection{Achievable Rate}
\label{subsec:achievable_rate}

To make the SIC analysis concrete, we focus on the two-user-per-branch configuration, and index the two co-branch users by $i\in\{0,1\}$. Following the cascaded-channel notation of~\cite{li_ris_otfs_noma,saglam_ris_miso_pda}, we collect the branch-$b$ STAR-RIS coefficients in a vector $\phi_b\in\mathbb{C}^{L}$, and define the cascaded BS-STAR-RIS-user equivalent channel of user $i$ on branch $b$ as $\mathbf{D}_{b,i} \triangleq \mathrm{diag}\bigl(h_{2,b,i}^{\rm H}\bigr)\,\mathbf{H}_1 \in \mathbb{C}^{L\times M}$, such that $\|\phi_b^{\rm T}\mathbf{D}_{b,i}\|^2$ captures the end-to-end effective channel gain after RIS steering. The per-branch NOMA power fractions are written in local form as $\eta_{b,i}\triangleq\eta_{u_{b,i}}$, where $u_{b,i}$ is the global index of the co-branch user with SIC rank $i$; $i=0$ denotes the weak user (lower $\gamma_u$) and $i=1$ the strong user (higher $\gamma_u$), consistent with the ordering established in Section~\ref{subsec:beamforming_decoding}.

For SIC to be valid at the strong user, the weak user's symbol must be correctly decoded both by itself and, subsequently, by the strong user. The two associated rates are
\begin{align}
R_{b,0\to0}&=\log_2\!\left(1+\frac{\eta_{b,0}P_{\rm t}\|\phi_b^{\rm T}\mathbf{D}_{b,0}\|^2}{\eta_{b,1}P_{\rm t}\|\phi_b^{\rm T}\mathbf{D}_{b,0}\|^2+\sigma_{b,0}^2}\right),\label{eq:r00}\\
R_{b,0\to1}&=\log_2\!\left(1+\frac{\eta_{b,0}P_{\rm t}\|\phi_b^{\rm T}\mathbf{D}_{b,1}\|^2}{\eta_{b,1}P_{\rm t}\|\phi_b^{\rm T}\mathbf{D}_{b,1}\|^2+\sigma_{b,1}^2}\right).\label{eq:r01}
\end{align}
As both conditions must hold, the achievable rate of user~0 is the minimum of the two,
\begin{equation}
R_{b,0}=\min(R_{b,0\to0},R_{b,0\to1})\geq R_{b,0,\min},
\label{eq:r0_min}
\end{equation}
where $R_{b,0,\min}$ denotes the minimum rate required to satisfy user~0's QoS constraint. Once user~0's contribution has been removed via SIC, user~1 decodes its own symbol free of intra-branch interference, and its achievable rate reduces to
\begin{equation}
R_{b,1}=\log_2\!\left(1+\frac{\eta_{b,1}P_{\rm t}\|\phi_b^{\rm T}\mathbf{D}_{b,1}\|^2}{\sigma_{b,1}^2}\right).
\label{eq:r1}
\end{equation}
Summing \eqref{eq:r0_min} and \eqref{eq:r1} over both branches yields the system sum-rate,
\begin{equation}
R_\Sigma=\sum_{b\in\{r,t\}}\left(R_{b,0}+R_{b,1}\right).
\label{eq:sum_rate}
\end{equation}
Throughout this work, the NOMA weights are fixed to $\eta_{b,i} = 1/K$ for all $b$ and $i$.

%%================================================================%%

\section{Problem Formulation}\label{sec:problem_formulation}
Our goal is to maximize the STAR-RIS-aided OTFS-NOMA DL sum-rate $R_\Sigma$ under user-QoS, STAR-RIS energy-conservation, and BS-power constraints. Since $g_u$ and $\eta_u$ are fixed by design (Section~\ref{subsec:beamforming_decoding}), the remaining decision variables are the STAR-RIS phase-shift coefficients $\phi_{\rm r}^l$ and $\phi_{\rm t}^l$, yielding
\begin{equation}
\begin{aligned}
\text{(P1):}\ \ &\underset{\phi_{\rm r}^l,\phi_{\rm t}^l}{\text{maximize}}\ \ R_\Sigma(\phi_{\rm r}^l,\phi_{\rm t}^l)\\
\text{s.t.}\ \ &\text{C1:}\ R_{b,i}\geq R_{b,i,\min},\ \forall b\in\{\rm r,\rm t\},\,i\in\{0,1\};\\
&\text{C2:}\ \beta_{\rm r}(l)+\beta_{\rm t}(l)=1,\ \beta_{\rm r}(l),\beta_{\rm t}(l)\in[0,1],\ \forall l;\\
&\text{C3:}\ \theta_{\rm r}(l),\theta_{\rm t}(l)\in[0,2\pi),\ \forall l=1,\dots,L;\\
&\text{C4:}\ \textstyle\sum_{u:\,\mathrm{branch}(u)=b}\eta_u \leq 1,\ \forall b\in\{\rm r,\rm t\};\\
&\text{C5:}\ \mathrm{tr}(\mathbf{G}\mathbf{G}^{\rm H})\leq P_{\rm t},\quad \mathbf{G}=[g_1,\dots,g_K].
\end{aligned}
\label{eq:P1}
\end{equation}

C1 guarantees each user's QoS. C2 enforces the STAR-RIS lossless energy conservation established in \eqref{eq:energy_conservation}, tying the reflection and transmission amplitudes at every element to a shared power budget. C3 restricts the STAR-RIS phase response of each branch to the unit circle. C4 states the per-branch NOMA power-budget feasibility for the fixed allocation $\eta_u = 1/K$; no fairness-based power ordering is imposed, since the power allocation is not optimized in this work. C5 is the BS transmit-power budget, satisfied automatically by the unit-norm MRT construction of $g_u$ in Section~\ref{subsec:beamforming_decoding} and the power-scaling in \eqref{eq:heisenberg}.

Problem \eqref{eq:P1} is non-convex. C2 couples the reflection and transmission amplitudes across every element, C3 confines the phases to a non-convex unit-modulus set, and the objective ties $\phi_{\rm r}^l$ and $\phi_{\rm t}^l$ through the cascaded channels $\mathbf{D}_{b,i}$ in a non-concave manner. The difficulty is compounded by the OTFS mobility model of Section~\ref{subsec:cascaded_channel}. While $\mathbf{H}_1$ is quasi-static, each $h_{b,i}(\tau,\nu)$ evolves with the delay-Doppler dynamics of its mobile user, such that the optimal solution of \eqref{eq:P1} changes every coherence interval and a fresh instance must, in principle, be re-solved at every time step. A classical AO scheme would alternate between updating the phase pair $(\theta_{\rm r},\theta_{\rm t})$ for fixed amplitudes $(\beta_{\rm r},\beta_{\rm t})$ and updating the amplitudes for fixed phases, iterating until convergence, and would repeat this procedure at every coherence interval. That is computationally prohibitive under OTFS mobility, which motivates the learning-based solution developed in Section~\ref{sec:proposed_solution}.

%%===================================================================%%

\section{Proposed Solution}
\label{sec:proposed_solution}

We solve \eqref{eq:P1} with a deep reinforcement learning agent that learns, offline, a direct mapping from the observed channel and system state to a feasible, near-optimal STAR-RIS configuration $\{\phi_{\rm r}^l,\phi_{\rm t}^l,\beta_{\rm r},\beta_{\rm t}\}$, and applies that mapping online in a single forward pass. This bypasses the per-coherence-interval re-optimization that AO would require. The agent is built on the SAC algorithm, augmented with a dedicated BSE sub-network that handles the coupled reflection/transmission energy-splitting trade-off imposed by C2.

\subsection{Markov Decision Process Formulation}
\label{subsec:mdp}
The joint design is cast as a Markov Decision Process $(\mathcal{S},\mathcal{A},\mathcal{R},\mathcal{P})$, with state, action, and reward defined as follows.
\begin{itemize}
    \item \textbf{State $s_t$} concatenates the previous action $a_{t-1}$, the current transmit and received power levels, and the observed cascaded channels $\{\mathbf{D}_{b,i}\}$ subject to one coherence block of feedback delay. The state layout also admits a channel-estimation noise term, although this term is not exercised in the reported experiments.

    \item \textbf{Action $a_t$} encodes the STAR-RIS phase-shift vectors $\theta_{\rm r},\theta_{\rm t}\in\mathbb{R}^{L}$ and the amplitude coefficients $\beta_{\rm r},\beta_{\rm t}\in[0,1]^{L}$, so that a single action fully specifies a candidate solution to \eqref{eq:P1} at time $t$. The beamformer $\mathbf{G}$ and the NOMA weights $\eta_u$ are excluded from the action space and are handled as described in Section~\ref{subsec:beamforming_decoding}.

    \item \textbf{Reward $r_t$} is the instantaneous NOMA sum-rate $R_\Sigma$ evaluated under the environment's true SIC decoding order, so that maximizing the expected cumulative reward is equivalent to solving \eqref{eq:P1} in expectation over the channel process.
\end{itemize}

\subsection{Soft Actor-Critic Backbone}
\label{subsec:sac_backbone}
SAC is an off-policy, maximum-entropy DRL algorithm that maximizes a trade-off between expected return and policy entropy,
\begin{equation}
J(\pi)=\mathbb{E}_\pi\!\left[\sum_t r(s_t,a_t)+\alpha\,\mathcal{H}(\pi(\cdot|s_t))\right],
\label{eq:sac_objective}
\end{equation}
where $\mathcal{H}(\pi(\cdot|s_t))$ is the policy entropy at state $s_t$ and $\alpha$ is a temperature coefficient balancing exploitation of the current best policy against continued exploration. SAC is well matched to \eqref{eq:P1} for two reasons. First, the OTFS channel changes every coherence interval, and the entropy term prevents the policy from collapsing onto a single configuration that would quickly become stale. Second, the STAR-RIS phase variables $\theta_{\rm r},\theta_{\rm t}\in\mathbb{R}^L$ are continuous, and SAC's Gaussian policy handles them without the discretization required by value-based methods such as DQN.

\subsection{Network Architecture}\label{subsec:architecture}
The agent comprises three sub-networks, namely a Gaussian policy network (actor), twin Q-networks (critics), and the BSE.
\begin{itemize}
\item \textbf{Actor:} The actor generates the raw STAR-RIS phase shifts $\phi_{\rm r}^l,\phi_{\rm t}^l$ from the current state. The beamforming matrix $\mathbf{G}$ is not produced by the actor; it is computed separately in closed form via MRT, as described in Section~\ref{sec:problem_formulation}.
\item \textbf{Twin critics:} Two independently trained Q-networks estimate $Q(s_t,a_t)$, and their minimum is taken to form the SAC target. This double-Q construction mitigates the value overestimation that would otherwise destabilize training under a fast time-varying channel.
\item \textbf{BSE:} A dedicated sub-network maps the current state to the element-wise reflection/transmission amplitude coefficients $\beta_{\rm r}(l),\beta_{\rm t}(l)$, resolving the coupled trade-off in C2 separately from the phase design carried out by the actor.
\end{itemize}

\subsection{Environment Interaction and Training}
\label{subsec:training}
At each step $t$, the actor supplies the STAR-RIS phase shifts and the BSE supplies the energy-splitting coefficients, and the beamformer $\mathbf{G}$ is assembled from the estimated effective channel via the closed-form MRT rule of Section~\ref{subsec:beamforming_decoding}. The environment advances the OTFS delay-Doppler channel by one coherence interval to reflect user mobility, evaluates the true NOMA-SIC sum-rate as the scalar reward $r_t$, and stores the resulting transition in the replay buffer. The critics are updated toward the SAC target, and their gradients drive the actor and the BSE, so that phase configuration and energy split are learned jointly under the same time-varying channel conditions encountered at deployment.

Algorithm~\ref{alg:star_ris_sac} summarizes the resulting training loop, in which one interaction step evaluates the current policy under the true SIC decoding order and one gradient step refines the actor, the critics, and the BSE toward the SAC objective in \eqref{eq:sac_objective}.
\begin{algorithm}[H]
\small
\caption{SAC-BSE Training Loop for STAR-RIS-Aided OTFS-NOMA}
\label{alg:star_ris_sac}
\begin{algorithmic}[1]
\STATE Initialize actor, twin critics (and their targets), Beta-Space Explorer, replay buffer, and channel feedback buffer.
\FOR{each time step}
\STATE Build the current state from the last action, power levels, and the latest delayed channel estimate.
\STATE Actor outputs STAR-RIS phase shifts; BSE outputs the reflection/transmission energy split.
\STATE Compute each user's beamforming vector from the estimated channel using MRT.
\STATE Advance the channel by one coherence interval and update the feedback buffer.
\STATE Order users on each STAR-RIS branch by channel gain, apply SIC, and compute each user's rate (used to form the reward).
\STATE Set the reward to the sum of user rates; store the transition in the replay buffer.
\IF{enough transitions are stored}
\STATE Sample a mini-batch and update the critics toward the SAC target.
\STATE Update the actor and the BSE to increase expected reward and policy entropy.
\STATE Softly update the target critics.
\ENDIF
\STATE Decay the exploration scale.
\ENDFOR
\end{algorithmic}
\end{algorithm}

\subsection{Computational Complexity}
\label{subsec:complexity}
For the trained agent produced by Algorithm~\ref{alg:star_ris_sac}, a full configuration ${\phi_{\rm r}^l, \phi_{\rm t}^l, \beta_{\rm r}, \beta_{\rm t}, \mathbf{G}}$ is obtained in a single forward pass through the actor and BSE networks, incurring an online complexity of $\mathcal{O}(L^2 K^2)$. In contrast, conventional AO using semidefinite relaxation (SDR) scales iteratively as $\mathcal{O}(I_{\mathrm{AO}}\cdot L^{3.5})$, where $I_{\mathrm{AO}}$ is the number of iterations required to converge. The dominant cost is thus shifted from online operation, where AO must re-solve \eqref{eq:P1} at every coherence interval, to offline training, which makes the proposed approach substantially more suitable for real-time deployment under OTFS mobility.

%%=========================================================%%
% \vspace{-3mm}
\section{Numerical Analysis}
\label{sec:numerical_analysis}
This section evaluates the proposed SAC-BSE agent against OTFS-only, NOMA-only, STAR-RIS-only, and fixed energy-splitting benchmarks, using the parameters listed in Table~\ref{tab:simulation_parameters}. The reported configuration serves two users on each STAR-RIS branch ($K_{\rm r}=K_{\rm }=2$), so that intra-branch NOMA multiplexing is exercised on both the reflection and transmission sides. Unless stated otherwise, the transmit power $P_{\rm t}$ is swept from 10 to 30~dBm and the number of STAR-RIS elements $L$ is swept over $\{4,8,16,32,64\}$.

\begin{table}[!t]
\centering
\caption{Simulation Setup and Network Hyperparameters}
\label{tab:simulation_parameters}
\renewcommand{\arraystretch}{1.15}
\begin{tabular}{|l|c|}
\hline
\textbf{Parameter} & \textbf{Value} \\
\hline

\multicolumn{2}{|l|}{\textit{System Topology and Setup}} \\
\hline
BS antennas ($M$) & 2 \\
\hline
STAR-RIS elements ($L$) & 16 \\
\hline
Users ($K$) & 4 \\
\hline
STAR-RIS user split ($K_{\rm r},K_{\rm t}$) & 2, 2 \\
\hline
Transmit power budget ($P_{\rm t}$) & 30 dBm \\
\hline
AWGN noise variance ($\sigma_w^2$) & $10^{-2}$ \\
\hline
NOMA power allocation ($\eta_u$) & $1/K = 0.25$ \\
\hline

\multicolumn{2}{|l|}{\textit{OTFS Channel Configuration (RIS--User Link)}} \\
\hline
Delay bins ($M_c$) & 16 \\
\hline
Doppler bins ($N$) & 8 \\
\hline
Propagation paths & 3 \\
\hline
Max. delay tap index & 4 \\
\hline
Max. Doppler tap magnitude & 2 \\
\hline
Clarke's SOS components & 8 \\
\hline
Fast-branch Doppler speed scale & 1.0--10.0 \\
\hline

\multicolumn{2}{|l|}{\textit{CSI Impairments}} \\
\hline
Feedback delay ($d$) & 1 coherence block \\
\hline
CSI noise std. ($\sigma_{\mathrm{csi}}$) & 0.0 \\
\hline

\multicolumn{2}{|l|}{\textit{STAR-RIS Amplitude Model}} \\
\hline
Minimum reflection coeff. ($\beta_{\min}$) & 0.6 \\
\hline

\multicolumn{2}{|l|}{\textit{DRL Agent (SAC + Beta-Space)}} \\
\hline
Discount factor ($\gamma$) & 0.99 \\
\hline
Soft update rate ($\tau$) & $10^{-3}$ \\
\hline
Entropy coefficient ($\alpha$) & 0.2 \\
\hline
Hidden dimensions & 256 \\
\hline
Learning rate & $3\times10^{-4}$ \\
\hline
Replay buffer size & $10^5$ \\
\hline
Minibatch size ($B$) & 256 \\
\hline
Initial exploration scale ($\lambda_0$) & 0.3 \\
\hline
Steps per episode & 100 \\
\hline

\end{tabular}
\end{table}
% \vspace{-3mm}
\subsection{Convergence of the Proposed Agent}
\label{subsec:convergence}

\begin{figure}[tb]
\centering
\includegraphics[scale=0.35]{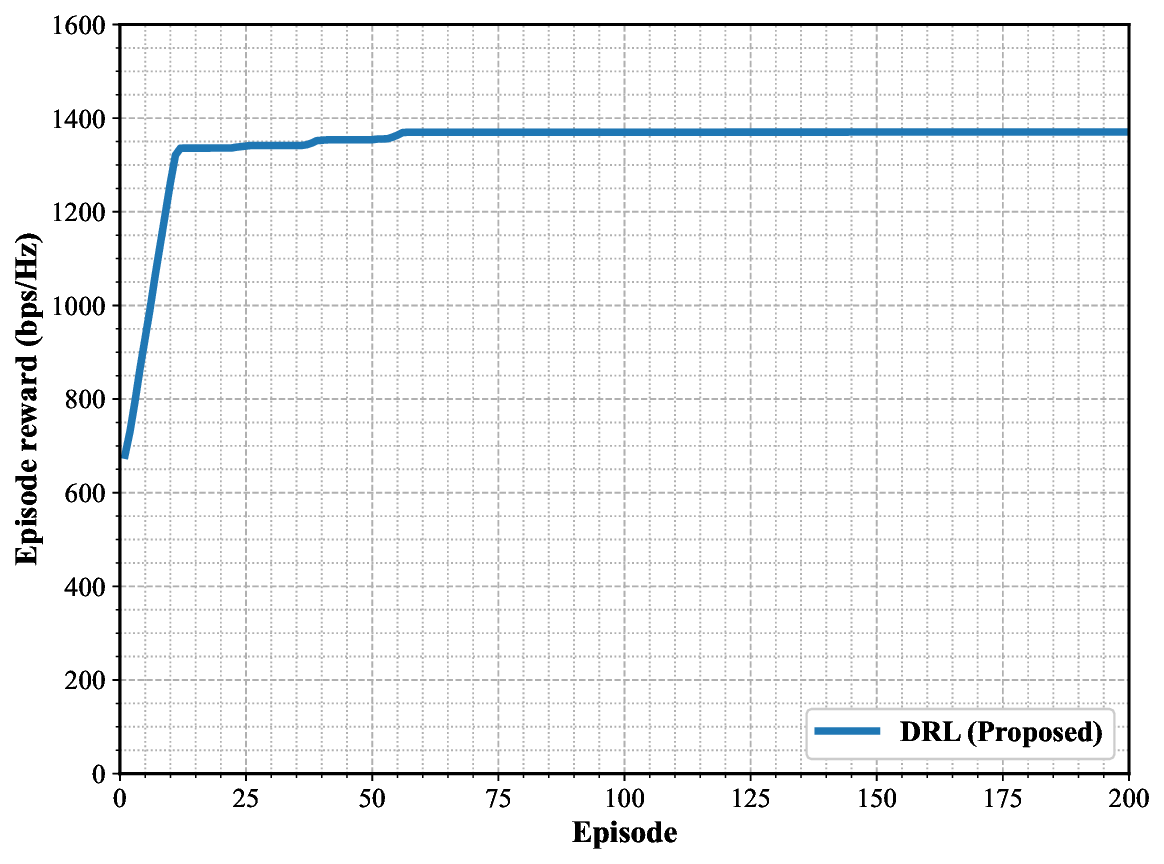}
\caption{Training convergence of the proposed SAC-BSE agent.}
\label{fig:conv}
\end{figure}

Fig.~\ref{fig:conv} reports the learning curve of the proposed SAC-BSE agent. The episode reward, equal to the instantaneous NOMA sum-rate achieved by the current policy, rises sharply over the first tens of episodes and then plateaus at a stable value, with no divergence or oscillation over the remaining training episodes. This behaviour indicates that the twin-critic architecture and the entropy-regularized SAC objective in \eqref{eq:sac_objective} are sufficient to stabilize training under the time-varying OTFS channel, and that the BSE sub-network learns an energy-splitting policy that is consistent with, rather than competing against, the actor's phase-shift policy.

% \vspace{-2mm}
\subsection{Robustness to User Mobility}\label{subsec:mobility}
Table~\ref{tab:mobility} sweeps the Doppler scale of the transmission-side, high-mobility users and reports the resulting mean sum-rate. In the lower portion of the range (Doppler scale $\leq 4$), the sum-rate remains within $0.2$ bps/Hz of its baseline value, confirming that the OTFS delay-Doppler representation combined with per-coherence-interval reconfiguration by the agent absorbs moderate Doppler spread. As the Doppler scale continues to rise, the sum-rate degrades gradually, dropping to $29.7$ bps/Hz at scale $64$, a total loss of about $10.5\%$ over a $128\times$ speed range. The absence of a sharp cliff, and the residual gap that even OTFS cannot fully close at very high Doppler, together indicate that the proposed design remains usable across the tested mobility range.
\begin{table}[tb]
\centering
\caption{Mean Sum-Rate versus User Speed (Doppler Scale)}
\label{tab:mobility}
\renewcommand{\arraystretch}{1.25}
\setlength{\tabcolsep}{4pt}
\begin{tabular}{|l|c|c|c|c|c|c|c|c|}
\hline
\textbf{Doppler Scale}       & 0.5  & 1    & 2    & 4    & 8    & 16   & 32   & 64   \\
\hline
\textbf{Sum-Rate (bps/Hz)}   & 33.2 & 33.2 & 33.1 & 33.0 & 32.6 & 30.9 & 29.9 & 29.7 \\
\hline
\end{tabular}
\end{table}

\subsection{Performance Comparison of Proposed Scheme over STAR-RIS, OTFS, and NOMA}
\label{subsec:benchmark}

\begin{figure}[tb]
\centering
\includegraphics[scale=0.35]{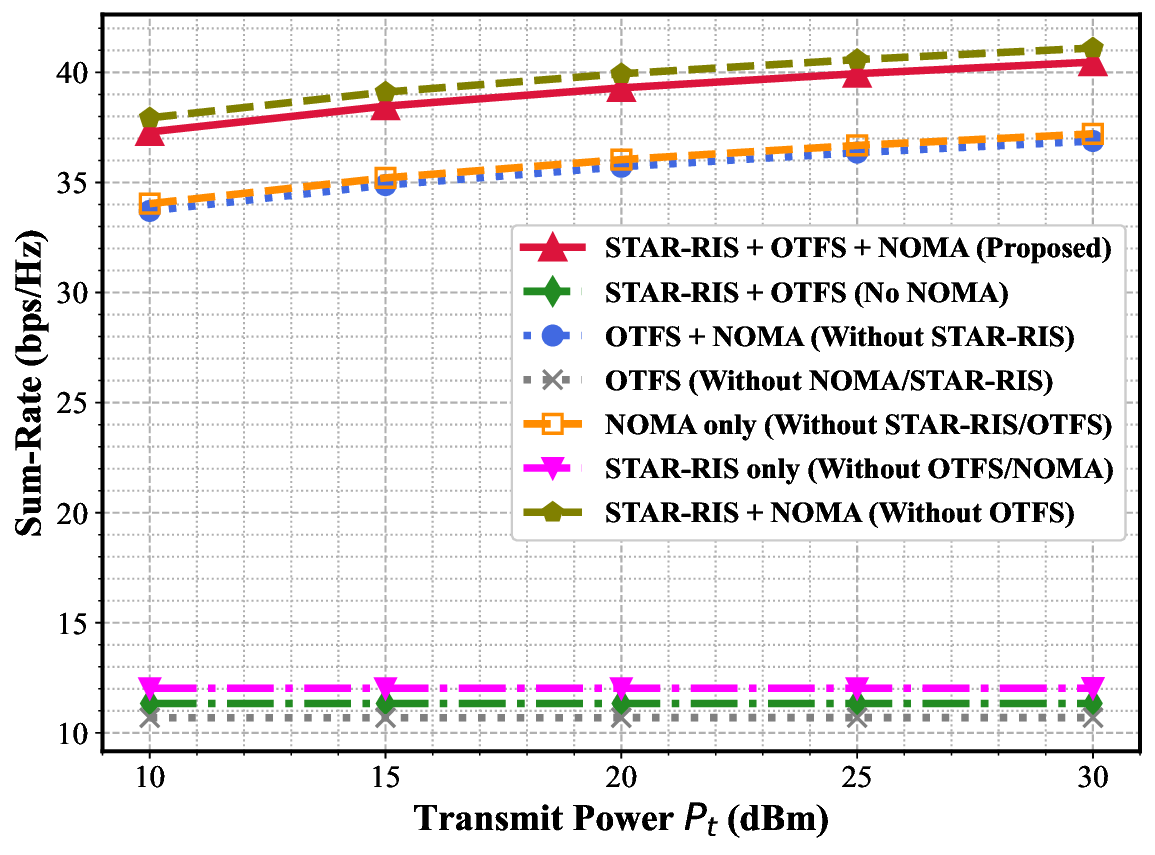}
\caption{Sum-rate versus $P_{\rm t}$ under single- and dual-technology ablations of the proposed design.}
\label{fig:benchmark}
\end{figure}

Fig.~\ref{fig:benchmark} shows the sum-rate as a function of the BS transmit power $P_{\rm t}$ for the proposed scheme and for five ablation benchmarks, each removing one of the three technologies. The proposed STAR-RIS+OTFS+NOMA scheme and the STAR-RIS+NOMA (without OTFS) benchmark achieve the two highest sum-rates across the entire power range, with STAR-RIS+NOMA slightly exceeding the proposed scheme. Under the static per-slot channel conditions used in this comparison, this indicates that OTFS incurs a small sum-rate cost relative to a scheme without it: OTFS trades a fraction of the peak spectral efficiency for the delay-Doppler robustness that Table~\ref{tab:mobility} shows is required once user mobility is introduced. Schemes lacking a STAR-RIS (OTFS-only, NOMA-only, and OTFS+NOMA without STAR-RIS) plateau at a markedly lower sum-rate that barely grows with $P_{\rm t}$, since without the surface providing an effective link, additional transmit power cannot be translated into a stronger received signal at either user group. STAR-RIS, rather than NOMA or OTFS individually, therefore emerges as the dominant contributor to the coverage gain in this topology.

\subsection{Effect of the Number of STAR-RIS Elements}
\label{subsec:ris_elements}

\begin{figure}[!t]
\centering
\includegraphics[scale=0.35]{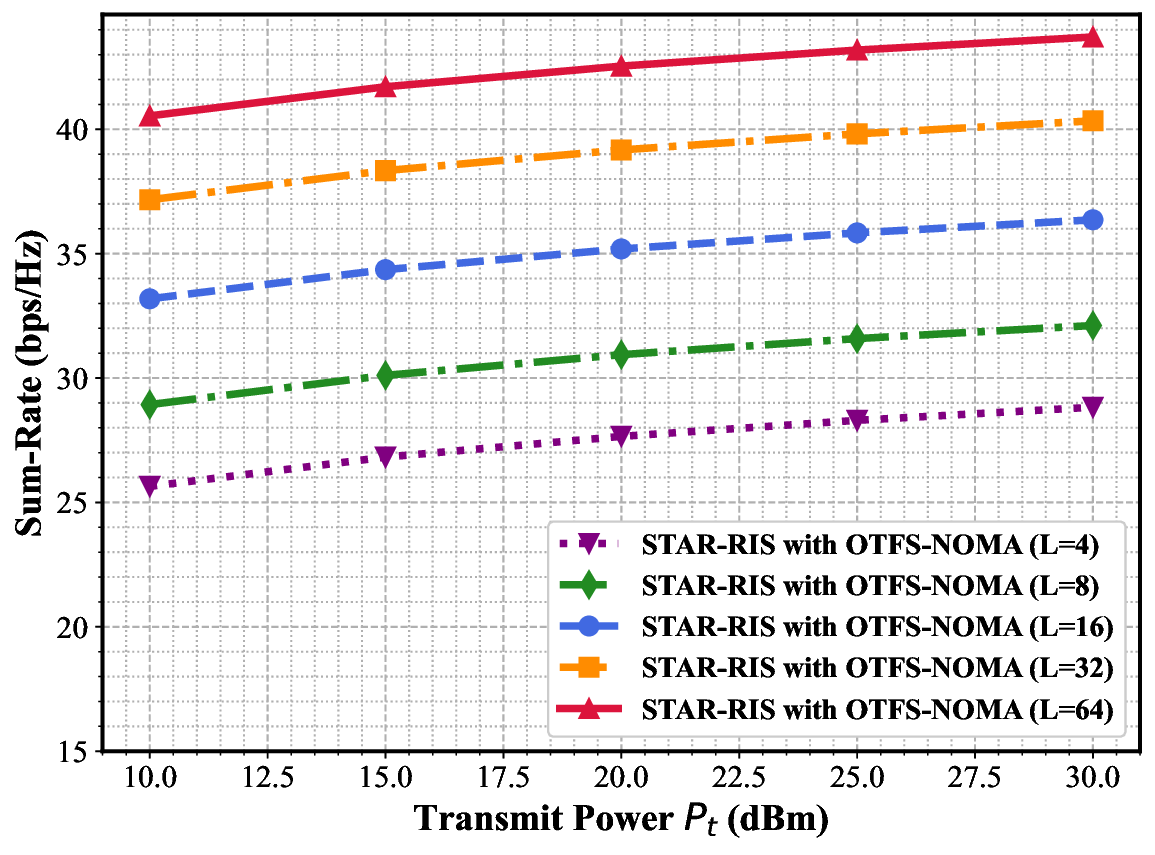}
\caption{Sum-rate versus $P_{\rm t}$ for varying STAR-RIS size $L$.}
\label{fig:ris_elements}
\end{figure}

Fig.~\ref{fig:ris_elements} repeats the transmit-power sweep for the proposed scheme as $L$ is increased. Larger surfaces achieve a higher sum-rate at every transmit-power level, and the gain from doubling $L$ is largest at small $L$ (for example from $L=4$ to $L=8$) and narrows at large $L$ (from $L=32$ to $L=64$), following the diminishing-return pattern characteristic of passive beamforming gain with array size. This confirms that the SAC-BSE agent continues to exploit the passive beamforming gain offered by a larger STAR-RIS, rather than saturating at a fixed action-space dimensionality, which would be a concern for a naively discretized DRL baseline.

\subsection{Impact of BSE}
\label{subsec:energy_split}

\begin{figure}[tb]
\centering
\includegraphics[scale=0.35]{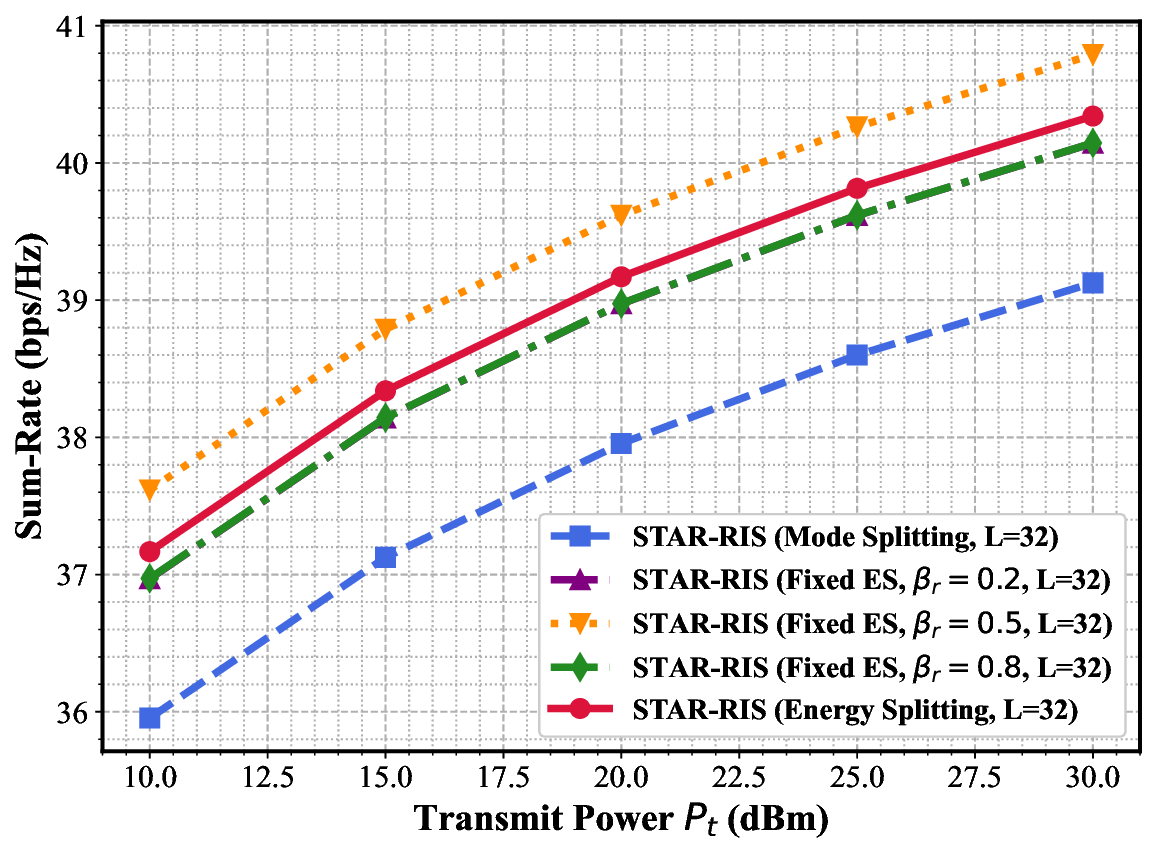}
\caption{Learned ES policy compared to MS and fixed-ratio ES baselines ($\beta_{\rm t} \in {0.2, 0.5, 0.8}$) with fixed $L = 32$.}
\label{fig:energy_split}
\end{figure}

Fig.~\ref{fig:energy_split} isolates the contribution of the BSE by comparing its learned, state-dependent energy-splitting policy against mode-switching STAR-RIS operation and three fixed energy-splitting ratios. The learned ES policy matches or exceeds every fixed-ratio baseline across the transmit-power range, confirming that no single fixed split is uniformly optimal and that letting the BSE adapt the split to the instantaneous channel state yields a real gain rather than an artifact of a favourably chosen fixed ratio. Mode switching, which restricts each element to either fully reflect or fully transmit, consistently underperforms both the learned ES policy and the best fixed-ratio baseline, which illustrates the benefit of the finer-grained energy-splitting mode assumed throughout the paper.
%%===========================================================%%
% \vspace{-2mm}
\section{Conclusion}
\label{sec:conclusion}

The paper addressed the sum-rate maximization of a STAR-RIS-aided OTFS-NOMA DL under user QoS, lossless energy conservation, and BS transmit-power constraints. In this context, we proposed SAC-BSE, a maximum-entropy DRL agent that delegated the two coupled design decisions to complementary functional roles. Learning both the phase shifts and the reflection transmission energy trade-off jointly under delayed CSI enabled the framework to align the trained policy with its intended deployment regime, and shifted the dominant computational burden from online solving to offline training. Numerical results confirmed that the resulting agent scaled its response with surface size, retained its operating point under increasing user mobility, and generalized across benchmark configurations that isolated the contribution of the surface, the modulation, and the multiple-access scheme. 
%These characteristics made the framework suitable for latency-constrained wireless deployments serving concurrent low- and high-mobility users through a shared programmable surface. 

% Incorporating explicit CSI noise into training, jointly learning the NOMA and beamforming layers, and extending the formulation to multi-cell coordination could be future extensions of the paper.

%%%%%%%%%%%%%%%%%%%%%%%%%%%%%%%%%%%
	% \vspace{-2mm}
	\balance
	\bibliographystyle{IEEEtran}
	\bibliography{references}

\end{document}